\documentclass[10pt]{blois07}

\usepackage{citesort}
\usepackage{amsmath}
\usepackage{graphicx}%
\usepackage{amsfonts}%
\usepackage{amssymb}
\usepackage{epsfig}

\newcommand{\bl}{\pmb{l}}
\newcommand{\bk}{\pmb{k}}

\newcommand{\be}{\begin{equation}}
\newcommand{\ee}{\end{equation}}

\newcommand{\bea}{\begin{eqnarray}}
\newcommand{\eea}{\end{eqnarray}}
\newcommand{\cM}{{\cal M}}
\usepackage{graphicx}
\usepackage{cite,./mcite}

\begin{document}
\title{Exclusive $J/\psi$ and $\Upsilon$ hadroproduction  as a probe of the QCD Odderon\footnote{$\;\;$Dedicated to the memory of Leszek {\L}ukaszuk, co-father of the odderon, who recently passed away.}}
\author{Lech Szymanowski\protect\footnote{$\;\;$talk presented at EDS07.}}
\institute{LPT, Universit\'e Paris XI, CNRS, Orsay, France, Universite de Li{\`ege}, Belgium \\ and  SINS Warsaw, Poland}
\maketitle
\begin{abstract}
We study  the exclusive production of $J/\psi$ or $\Upsilon$ in $pp$ and $\bar pp$ collisions, where the meson emerges 
from the pomeron--odderon and the pomeron--photon fusion. 
We estimate the cross sections for these processes
for the kinematical conditions of the Tevatron and of the LHC.
\end{abstract}

\section{Introduction}

The new analysis of exprimental data on the 
 exclusive hadroproduction processes by the CDF collaboration \cite{tevatron} shows that
 these types of processes can be  objects of detailed study at the Tevatron and in the near future at the LHC.
 Up to now, the most intensively   studied exclusive hadroproduction processes include the dijet or the $\gamma \gamma$ production in the central
 rapidity region and the Higgs meson production \cite{khose}, see Fig. 1. 
\begin{figure}[t]
\epsfxsize=3.5cm
\begin{center}
\epsffile{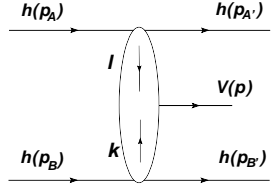}
\end{center}
\caption{\it Kinematics of the exclusive meson production in $pp$ ($p\bar p$) scattering.}
\end{figure} 
Here, we discuss the exclusive hadroproduction of  
$J/\Psi$  and  $\Upsilon$ mesons, i.e.
\be
pp\,(\bar{p})\,\rightarrow \, p' \, V \,p''\,(\bar{p}''\ )\;,\;\;\;\;\ \mbox{where}\;\;\;\;\; V=J/\psi,\, \Upsilon\;.
\label{genprocess}
\ee

\noindent
 The main motivation of our recent study \cite{bmsc} of the process (\ref{genprocess}) is that the
 production of a charmonium $V$, with
 the quantum numbers $J^{PC}=1^{-\ -}$, occurs as the result of a pomeron-odderon or pomeron-photon fusion. 
 Such studies can thus probe the dynamics of the odderon \cite{Lukaszuk}, i.e. 
 the counterpart with negative charge parity of the pomeron. Odderon escapes experimental verification 
 and until now has remained a mystery, although various ways to detect it through its interference with a pomeron mediated amplitude \cite{PomOdd} have been recently proposed  (for a review see \cite{Ewerz-review}).  
 
\section{The scattering amplitude}

In the lowest order of perturbative QCD, the pomeron and the odderon are described by the exchange of two and three non-interacting gluons, respectively. The lowest order contribution to the hadroproduction 
\begin{equation}
\label{process}
 h(p_A)+h(p_B)\to h(p_{A'}) + V(p) +h(p_{B'})
\end{equation}
is illustrated by diagrams of Fig. 2, from which the diagrams (a,b) describe the pomeron-odderon fusion and (c,d)  
the photon-pomeron fusion. The momenta of particles are parametrized by the Sudakov decompositions
\begin{equation}
\label{outpart}
p_{A'}=(1-x_{A})p_A + \frac{\bl^2}{s(1-x_A)}p_B -l_\perp\ , \;\;\;\;p_{B'}=\frac{\bk^2}{s(1-x_B)}p_A +  (1-x_B)p_B -k_\perp\;,
\end{equation}
with ${\bl}^2=-l_\perp \cdot l_\perp \approx - (p_A-p_{A'})^2\equiv -t_A$, $\;\;\;{\bk}^2=-k_\perp \cdot k_\perp \approx (p_B - p_{B'})^2 \equiv -t_B\;\;\;$ and 

$$
p = \alpha_p p_A + \beta_p p_B +p_\perp
$$
\begin{equation}
\alpha_p=x_A - \frac{\bk^2}{s(1-x_B)} \approx x_A \;,\;\;\beta_p=x_B-\frac{\bl^2}{s(1-x_A)}\approx x_B\;,\;\;\;p_\perp = l_\perp + k_\perp\;,
\end{equation}
which lead to the mass-shell condition for the vector meson, 
$V=J/\psi, \Upsilon$,\\
 $m_{V}^2 = s x_A x_B -(\bl +\bk)^2\;.$
\begin{figure}[t]
\begin{center}
{\large\bf a)}
\epsfxsize=2.8cm \epsffile{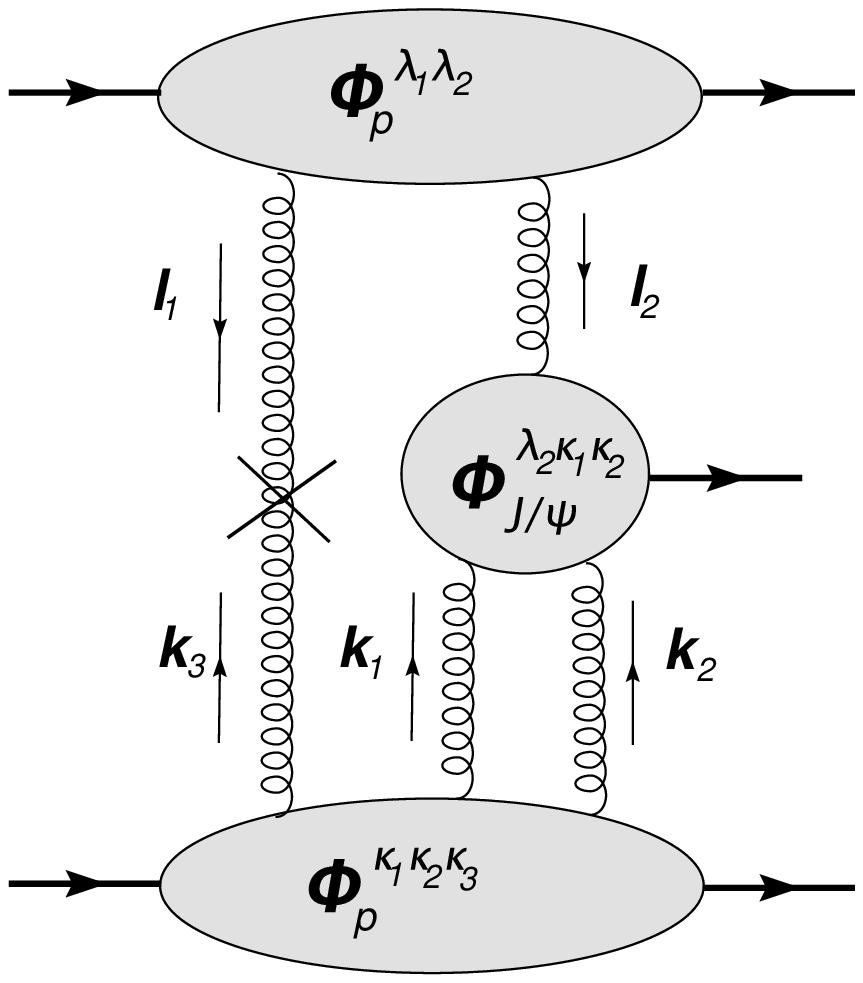} \hspace{0.3cm}
{\large\bf b)}
\epsfxsize=2.8cm \epsffile{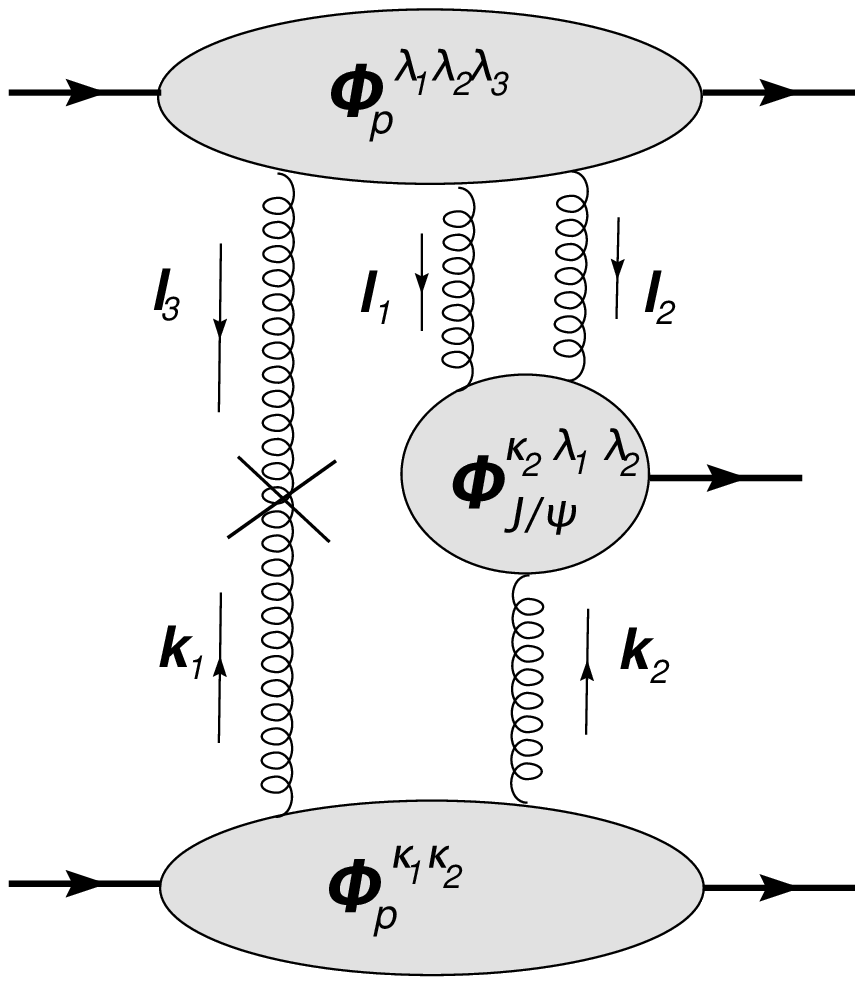} \hspace{0.3cm}
{\large\bf c)}
\epsfxsize=2.8cm \epsffile{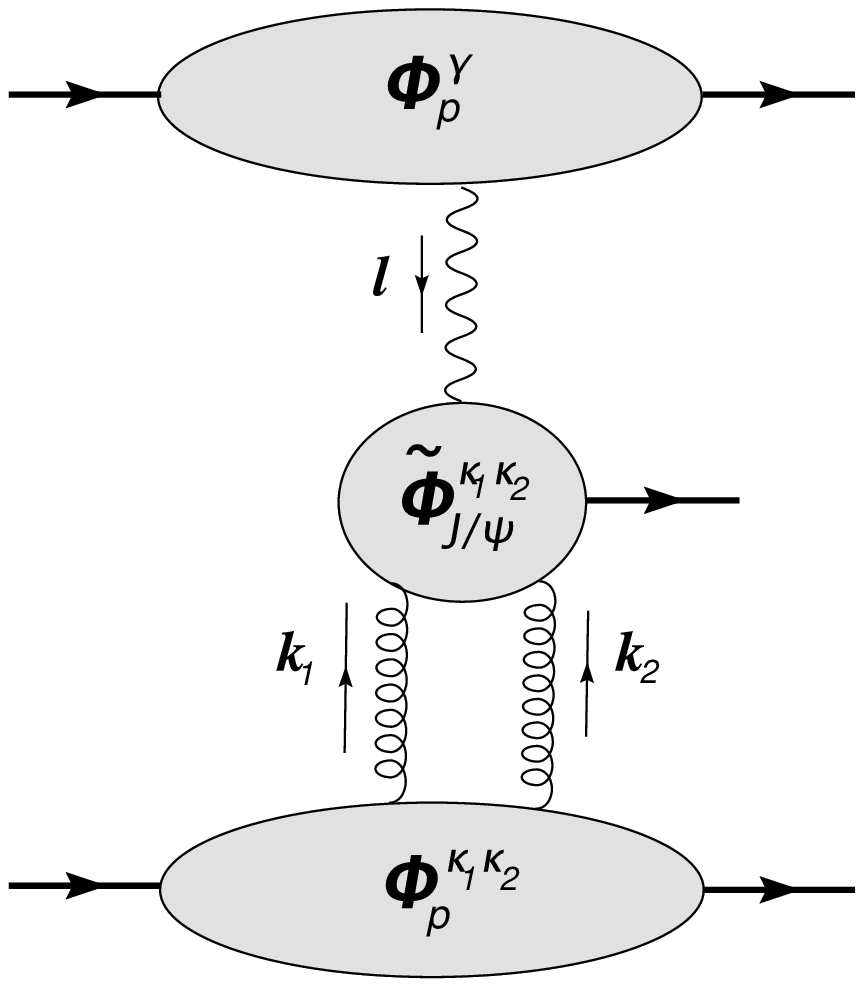} \hspace{0.3cm}
{\large\bf d)}
\epsfxsize=2.8cm \epsffile{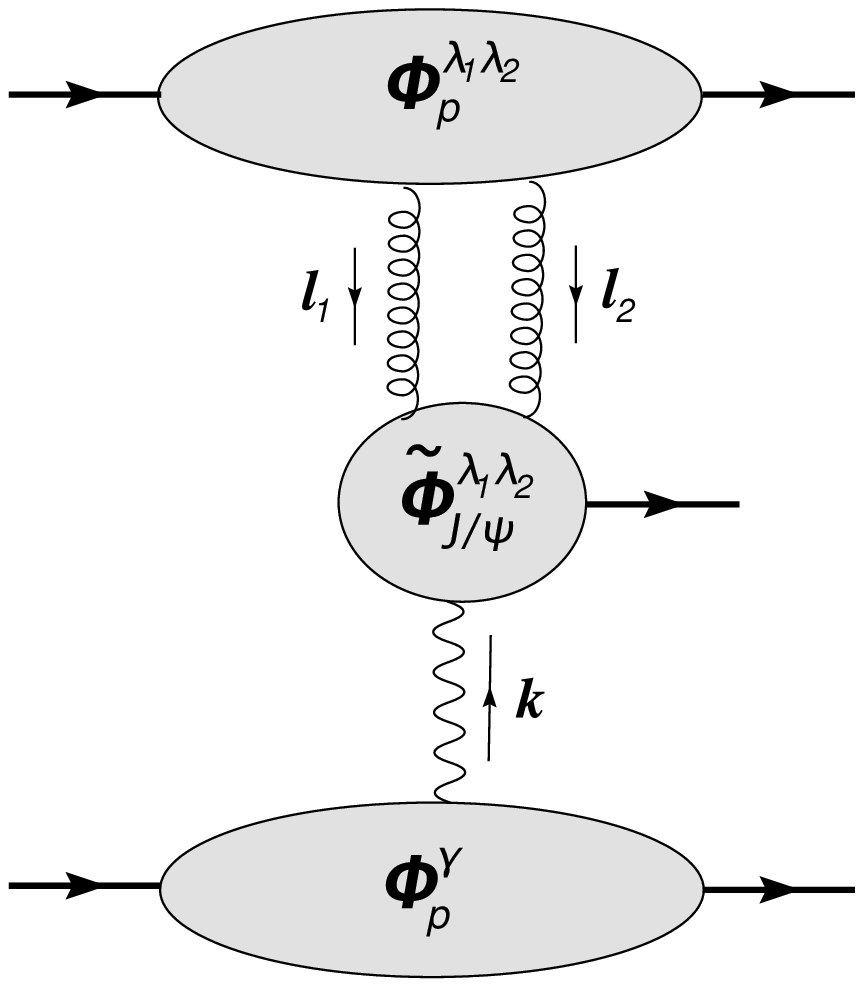} 
\end{center}
\caption{\it The lowest order diagrams contributiong to  the pomeron-odderon fusion (a,b) and  the pomeron--photon fusion (c,d)  for
 vector meson production.}
\end{figure}

\noindent
The scattering amplitude written within the $k_\perp$-factorization approach is a convolution in transverse momenta of
$t-$channel fields. For instance,  the contribution of Fig. 2a  reads:
\bea
\label{impactA}
&& \cM_{P\;O}=
 \\
&&-is\;\frac{2\cdot3}{2!\,3!}\,\frac{4}{(2\pi)^8}\,\int \frac{d^2 \bl_1}{\bl_1^2}\,\frac{d^2 \bl_2}{\bl_2^2}\ \delta^2(\bl_1 + \bl_2 - \bl)\,\frac{d^2 \bk_1}{\bk_1^2}\,\frac{d^2 \bk_2}{\bk_2^2}\,\frac{d^2 {\bk}_3}{\bk_3^2}\ \delta^2(\bk_1+\bk_2+\bk_3-\bk)
\nonumber \\
&&
\times \delta^2(\bk_3+\bl_1)\,\bk_3^2\;\delta^{\lambda_1 \kappa_3}
\cdot\Phi^{\lambda_1 \lambda_2}_P(\bl_1,\bl_2)\cdot\Phi^{\kappa_1 \kappa_2 \kappa_3}_P(\bk_1,\bk_2,\bk_3)\cdot
\Phi_{J/\psi}^{\lambda_2 \kappa_1 \kappa_2}(\bl_2, \bk_1,  \bk_2)
\nonumber \;,
\eea
where $\Phi^{\lambda_1 \lambda_2}_P(\bl_1,\bl_2)$ and $\Phi^{\kappa_1 \kappa_2 \kappa_3}_P(\bk_1,\bk_2,\bk_3)$ 
are the impact factors describing the coupling of the pomeron and the odderon to  scattered hadrons, respectively,
whereas
$\Phi_{J/\psi}^{\lambda_2 \kappa_1 \kappa_2}(\bl_2, \bk_1,  \bk_2)$ is the effective $J/\psi$-meson production vertex. 

\noindent
The proton impact factors are non-perturbative objects and we describe them within the Fukugita-Kwieci\'nski eikonal model \cite{Fukugita}. For the pomeron exchange the impact factor of the proton is the product
\be
\label{imfaPom}
\Phi^{\lambda_1 \lambda_2}_P(\bl_1,\bl_2)=3\ \Phi^{\lambda_1 \lambda_2}_q(\bl_1,\bl_2)\ 
{\cal F}_P(\bl_1,\bl_2)\;,
\ee
 of the impact factor of a quark
\be
\label{imfaPQ}
\Phi^{\lambda_1 \lambda_2}_q(\bl_1,\bl_2)= -\bar \alpha_s\cdot 8\pi^2 \cdot \frac{\delta^{\lambda_1 \lambda_2}}{2\, N_c}\;,
\ee
and the phenomenological form-factor ${\cal F}_P$ describing the proton internal structure
\be
\label{ffPom}
{\cal F}_P(\bl_1,\bl_2)= F(\bl_1+\bl_2,0,0)-F(\bl_1,\bl_2,0)\;,
\ee
which vanishes when any of $\bl_i \to 0$, as required by  colour gauge invariance.
The function $F(\bk_1,\bk_2,\bk_3)$ is chosen in the form 
\be
\label{F}
F(\bk_1,\bk_2,\bk_3)= \frac{A^2}{A^2 +\frac{1}{2}\left( (\bk_1-\bk_2)^2 + (\bk_2-\bk_3)^2 +(\bk_3-\bk_1)^2 \right)},
\ee
with $A$  a phenomenological constant chosen to be half of the 
$\rho$~meson mass, $A=m_\rho/2 \approx 384\,$MeV.
The analogous impact-factor for the odderon exchange reads
\be
\label{imfaOdd}
\Phi^{\kappa_1 \kappa_2  \kappa_3}_P(\bk_1,\bk_2,\bk_3)=3\  \Phi^{\kappa_1 \kappa_2 \kappa_3}_q(\bk_1,\bk_2,\bk_3)
{\cal F}_O(\bk_1,\bk_2,\bk_3)\;,
\ee
where
\be
\label{imfaOQ}
\Phi^{\kappa_1 \kappa_2 \kappa_3}_q(\bk_1,\bk_2,\bk_3) =i\ \bar \alpha_s^{\frac{3}{2}}\ 2^5\ \pi^\frac{7}{2}\ \frac{d^{\kappa_3 \kappa_2 \kappa_1}}{4N_c} \;,
\ee
and the form-factor ${\cal F}_O$ has a form
\be
\label{ffOdd}
\hspace*{-0.4cm}{\cal F}_O(\bk_1,\bk_2,\bk_3)=F(\bk=\bk_1+\bk_2+\bk_3,0,0) -\sum\limits_{i=1}^3\ F(\bk_i, \bk-\bk_i,0) +
2\ F(\bk_1,\bk_2,\bk_3)\;.
\ee
\noindent
The derivation of the effective production vertex of $J/\psi$, $\Phi_{J/\psi}^{\lambda_2 \kappa_1 \kappa_2}(\bl_2, \bk_1,  \bk_2)$, in Eq.~(\ref{impactA}) is one of the main results of our study. The charmonium is treated in the non-relativistic approximation and it is described by the $\bar c c \to J/\psi$ vertex
\be
\label{JPsivertex}  
\langle \bar c \ c |J/\psi \rangle = \frac{g_{J/\psi}}{2}\ \hat \varepsilon^{\ *}(p)\left( p \cdot \gamma  + m_{J/\psi}  \right)\;,\;\;\;\;
m_{J/\psi}=2m_c\;,
\ee
with the coupling constant $g_{J/\psi}$ related to the electronic width $\Gamma^{J/\psi}_{e^+e^-}$ of the $J/\psi \to e^+\ e^-$ decay
\be
\label{width}
g_{J/\psi}=\sqrt{ \frac{3m_{J/\psi} \Gamma^{J/\psi}_{e^+e^-}}{16\pi \alpha_{em}^2 Q_c^2} }\;,\;\;\;\;Q_c=\frac{2}{3}\;.
\ee 
The effective vertex $g+2g \to J/\psi$ is described by the sum of the contributions of the diagrams
in Fig.~3 which has the form 
\begin{figure}[t]
\centerline{
\epsfysize=4.0cm
\epsfxsize=9.0cm
\epsffile{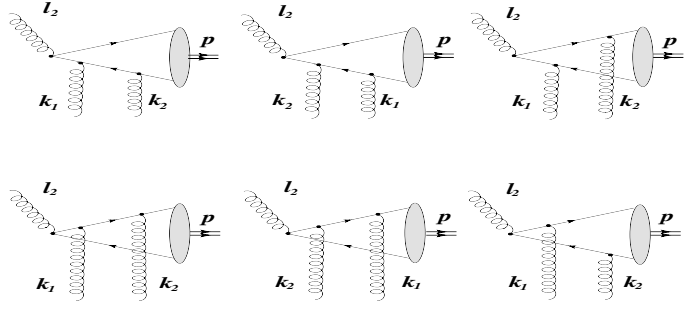} }
\caption{\it The six diagrams defining the effective vertex $g+2g \to J/\psi$.}
\end{figure}
\bea
\label{vertexPO}
&&\Phi_{J/\psi}^{\lambda_2 \kappa_1 \kappa_2}(\bl_2, \bk_1,  \bk_2) =
\alpha_s ^{3\over 2} \, 8\pi^{3\over 2}\; \frac{d^{\kappa_1\kappa_2\lambda_2}}{N_c}\;V_{J/\psi}(\bl_2, \bk_1,  \bk_2), \nonumber
\\
&& \hspace{-1cm}V_{J/\psi}(\bl_2, \bk_1,  \bk_2)=
\\
&&4\pi m_c g_{J/\psi}\left[ - \frac{x_B \varepsilon^*\cdot p_B + \varepsilon^*\cdot l_{2\perp}}{\bl_2^2+(\bk_1+\bk_2)^2+4m_c^2} + \frac{\varepsilon^*\cdot l_{2\perp} +\varepsilon^*\cdot p_B\left( x_B - \frac{4\bk_1\cdot \bk_2}{sx_A} \right)}{\bl_2^2+(\bk_1-\bk_2)^2 +4m_c^2}  \right]\;.
\nonumber
\eea
In the numerical analysis we set $\alpha_s(m_c)=0.38$ and $\alpha_s(m_b)=0.21$.

\noindent
The analogous formula which describes the photon-pomeron fusion in Fig.~2c has the form
\bea
\label{impactC}
&&{\cal M}_{\gamma\, P}=
\\
&&\hspace{-1cm}-\frac{1}{2!} \cdot s \cdot \frac{4}{(2\pi)^4\ \bl^2} \ \Phi^\gamma_P(\bl)\ \int \frac{d^2\bk_1}{\bk_1^2}
\frac{d^2 \bk_2}{\bk_2^2}\ \delta^2(\bk_1+\bk_2-\bk)\ \Phi^{\kappa_1\kappa_2}_P(\bk_1,\bk_2)\ \tilde \Phi^{\kappa_1\kappa_2}_{J/\psi}(\bl,\bk_1,\bk_2)\;,
\nonumber
\eea
where $\Phi^\gamma_P(\bl)$ is the phenomenological form-factor of the photon coupling to the proton chosen as
$\Phi^\gamma_P(\bl)= -i e \cdot F(\bl,0,0)$.
The pomeron impact factor
$\Phi^{\kappa_1\kappa_2}_P(\bk_1,\bk_2)$ is given by Eq.~(\ref{imfaPom})  and
$\tilde \Phi^{\kappa_1\kappa_2}_{J/\psi}(\bl,\bk_1,\bk_2)$ is the corresponding effective vertex
expressed through $V_{J/\psi}(\bl, \bk_1,  \bk_2)$ in Eq.~(\ref{vertexPO})
\be
\label{vertexGP}
\tilde \Phi^{\kappa_1\kappa_2}_{J/\psi}(\bl,\bk_1,\bk_2) = \, \alpha_s \,  eQ_c\, 8\pi\,
\frac{\delta^{\kappa_1\kappa_2}}{N_c}\,V_{J/\psi}(\bl, \bk_1,  \bk_2)\, .
\ee
The phases of the scattering amplitudes describing the two mechanisms of $J/\psi$-meson production differ by the factor $i=e^{i\pi/2}$. Consequently, they do not interfere and they contribute to the cross section as a sum of two independent contributions.

\section{Estimates for the cross sections}
We analyse the contributions of pomeron-odderon fusion and the photon-pomeron fusion separately. Denoting ${\cal M}_{PO} ^{\mathrm{tot}} = {\cal M}_{PO} +  {\cal M}_{OP}$
and ${\cal M}_{\gamma P} ^{\mathrm{tot}} = {\cal M}_{\gamma P} +  {\cal M}_{P \gamma}$ we calculate the differential cross sections with respect to the rapidity $y\approx \frac{1}{2}\log(x_A/x_B)$, the squared momentum transfers in the two $t$-channels, $t_A,\ t_B$, and the azimuthal angle $\phi$ between $\bk$ and $\bl$ 
\be
\label{dsdy}
{d \sigma^{(\varepsilon)}_i \over dy\, dt_A\, dt_B\, d\phi } = 
{1 \over 512\pi^4 s^2} \, |{\cal M}_{i}^{\mathrm{tot}}|^2\;\;\;\;\;\;i=PO,\ \gamma P\;, 
\ee
and the partially integrated cross sections
\be
{d \sigma_i \over dy} = \sum_{\varepsilon} \;
\int_{t^A _{\min}} ^{t_{\max}} \,dt_A\,\int_{t^B_{\min}} ^{t_{\max}} \,dt_B\,
\int_0 ^{2\pi} d\phi  
\, {d \sigma^{(\varepsilon)}_i \over dy\, dt_A\, dt_B\, d\phi}\; ,
\label{dsdt}
\ee 
with $t^A_{min}= 0=t^B_{min}$ for the $PO$-fusion and $t^A_{min}=m_p^2x_A^2$, $t^B_{min}=m_p^2x_B^2$ for the $\gamma P$-fusion, and we set $t_{max}=1.44\ $GeV$^2$. This leads to the naive predictions shown in the Table~1.
\begin{table}[t]
\label{tab1}
\begin{center}%
\begin{tabular}[c]{|c|c|c|c|c|}\hline\hline
$d\sigma/dy$ & 
\multicolumn{2}{c|}{$ J/\psi$} &
\multicolumn{2}{c|}{$\Upsilon$} \\ \cline{2-5}
 &  odderon & photon & odderon & photon    \\ \hline
$p\bar p$ & 20~nb & 1.6~nb &  36~pb & 1.1~pb    \\
$pp$      & 11~nb & 2.3~nb &  21~pb & 1.7~pb    \\\hline \hline
\end{tabular}
\end{center}
\caption{
 Na\"{i}ve cross sections $d\sigma /dy$ given by (\ref{dsdt}) 
for the exclusive $J/\psi$ and $\Upsilon$ production in $pp$ and $p\bar p$ 
collisions by the odderon-pomeron fusion, assuming $\bar\alpha_s=1$ and
analogous cross sections  $d\sigma_{\gamma} /dy$ for the photon 
contribution.  }
\end{table}
More realistic cross-sections are obtained by taking into account phenomenological improvements, such as related to the
BFKL evolution (which is very important for the pomeron exchange and which may be  omitted for the odderon exchange \cite{bmsc}), the effects of soft rescatterings of hadrons, and the precise determination of the value of the model parameter $\bar \alpha_s$ in the impact factors.  For that we write the corrected cross-sections in the form
\be
\hspace*{-1cm}
\left.
{d \sigma^{\mathrm{corr}}_{PO} \over dy} 
\right|_{y=0}
\, = \, 
\bar\alpha_s^5\, S_{\mathrm{gap}}^2\, E(s,m_V) \,{d \sigma_{PO} \over dy},\;\;\;\;\;\;\;\;\;\left. {d \sigma^{\mathrm{corr}}_{\gamma P} \over dy}\right|_{y=0} \, = \, 
\bar\alpha_s^2\, E(s,m_V) \,{d \sigma_{\gamma P} \over dy},
\label{master}
\ee
where ${d \sigma_{PO/\gamma P}/ dy}$ are the cross sections given by (\ref{dsdt}) at
$\bar\alpha_s=1$.
The BFKL evolution for pomeron exchange  is taken into account by inclusion of the enhancement factor, which for the central production (i.e. for the rapidity $y = 0$) has the form 
\be
E(s,m_V) = (x_0 \sqrt{s}/m_V)^{2\lambda}.
\ee
Here, $x_0$ is
the maximal fraction of incoming hadron momenta exchanged in the $t-$channels (or the initial condition for the BFKL evolution) and it is set $x_0=0.1$.
The effective pomeron intercept $\lambda$ is determined by HERA data and it equals $\lambda=0.2$
($\lambda=0.35$) for the $J/\psi$ ($\Upsilon$) production \cite{HERA}.

\noindent
The gap surviving factor $S_{\mathrm{gap}}^2$ for the exclusive production via the pomeron-odderon fusion is fixed by the results of Durham two channel eikonal model 
\cite{two-channel}: $S_{\mathrm{gap}}^2=0.05$ for the exclusive production at the Tevatron and $S_{\mathrm{gap}}^2=0.03$ for LHC. In the case of production from the photon-pomeron fusion, 
$S_{\mathrm{gap}}^2=1$ \cite{KMR-phot}.

\noindent
The available estimates of the effective strong coupling constant 
$\bar\alpha_s$ in the Fukugita--Kwieci\'{n}ski model 
yield results with rather large spread: from $\bar\alpha_s\approx 1$  \cite{Fukugita}, through $\bar\alpha_s\approx 0.6-0.7$ determined from the HERA data \cite{bmsc} to $\bar\alpha_s\approx 0.3$ determined from data on  elastic
$pp$ and $p\bar p$ scattering \cite{Ewerz-coupling}. 
This led us to introduce  three scenarios which differ by the values of $\bar\alpha_s$ and of $S_{\mathrm{gap}}^2$.
In the {\em optimistic scenario} we  use a large value of the
coupling, $\bar\alpha_s = 1$, combined with the gap survival factors 
obtained in the Durham two-channel eikonal model. We believe that the best estimates should follow 
from the {\em central scenario} defined by 
$\bar\alpha_s=0.75$, and Durham group estimates $S_{\mathrm{gap}}^2 =0.05$ ($S_{\mathrm{gap}}^2=0.03$) 
at the Tevatron (LHC). The {\em pessimistic scenario} is defined by $\bar\alpha_s=0.3$ and $S_{\mathrm{gap}}^2=1$.

\noindent
Table 2 shows our predictions for the phenomenologically improved cross sections in all three scenarios. 
Their magnitudes justify our hope that the process (\ref{genprocess}) is a subject of experimental 
study in the near future at the Tevatron and at the LHC \cite{alice}. The encouraging feature of our results is due to the fact, that
the measurement of the $t_i$ dependence of the cross section partially permits filtering out the $\gamma\ P$ contributions and to uncover  the $PO$ ones.
\begin{table}[t]
\begin{center}%
\label{tab2}
\begin{tabular}[c]{|c|c|c|c|c|}\hline\hline
$d\sigma^{\mathrm{corr}}/dy$ & 
\multicolumn{2}{c|}{$ J/\psi$} &
\multicolumn{2}{c|}{$\Upsilon$} \\ \cline{2-5}
 &  odderon & photon & odderon & photon \\ \hline
Tevatron &
0.3--1.3--5~nb & 
0.8--5--9~nb &  
0.7--4--15~pb & 
0.8--5--9~pb  \\
LHC      & 
0.3--0.9--4~nb & 
2.4--15--27~nb  & 
1.7--5--21~pb & 
5--31--55~pb\\\hline \hline
\end{tabular}
\end{center}
\caption{
 The phenomenologically corrected cross sections $d\sigma^{\mathrm{corr}} /dy|_{y=0}$ 
 for the exclusive $J/\psi$ and $\Upsilon$ production 
in $pp$ and $p\bar p$ collisions by the pomeron--odderon  
fusion, and analogous cross sections  
$d\sigma^{\mathrm{corr}} _{\gamma} /dy|_{y=0}$ for the photon contribution
 for the 
pessimistic--central--optimistic scenarios.}
\end{table}

\section*{Acknowledgments}

I acknowledge common research and discussions with A. Bzdak, J-R. Cudell and L.Motyka. 
This work is partly supported by the Polish (MEiN) research 
grant 1~P03B~028~28 and by the
Fonds National de la Recherche Scientifique (FNRS, Belgium).
I acknowledge a warm hospitality  at 
Ecole Polytechnique and at LPT-Orsay.

\begin{footnotesize}
\bibliographystyle{blois07} 
{\raggedright

\begin{thebibliography}{99}

\bibitem{tevatron} K. Goulianos, see talk at this conference.

\bibitem{khose} V. Khose, see talk at this conference.


\bibitem{bmsc} 
A.~Bzdak, L.~Motyka, L.~Szymanowski and J.~R.~Cudell,
  Phys.\ Rev.\  D {\bf 75}, 094023 (2007)
  [arXiv:hep-ph/0702134].




\bibitem {Lukaszuk}L. Lukaszuk and B. Nicolescu, Nuovo Cimento Letters
\textbf{8}, 405 (1973).


\bibitem{PomOdd} S.~J.~Brodsky et al.,
  Phys.\ Lett.  B {\bf 461}, 114 (1999) ; 
Ph.~Hagler et al.,  
  Phys.\ Lett.  B {\bf 535}, 117 (2002),\, Eur.\ Phys.\ J.  C {\bf 26},
261 (2002);
I.~F.~Ginzburg et al.,
  Eur.\ Phys.\ J.  C {\bf 30}, 002 (2003).



\bibitem {Ewerz-review}C. Ewerz, hep-ph/0306137.


\bibitem {Fukugita}M. Fukugita and J. Kwieci\'{n}ski, Phys.\ Lett.\ B
\textbf{83} (1979) 119; see also 
J.~F.~Gunion and D.~E.~Soper,
Phys.\ Rev.\ D {\bf 15} (1977) 2617 and 
J.~R.~Cudell and B.~U.~Nguyen,
  Nucl.\ Phys.\  B {\bf 420} (1994) 669.

\bibitem{HERA}
H1 Collaboration:  S.~Aid {\it et al.},
Nucl.\ Phys.\ B {\bf 472} (1996) 3;
H1 Collaboration: C.~Adloff {\it et al.},
Phys.\ Lett.\ B {\bf 483} (2000) 23;
%
H1 Collaboration: A.~Aktas {\it et al.},
Eur.\ Phys.\ J.\ C {\bf 46} (2006) 585.
  ZEUS Collaboration: J.~Breitweg {\it et al.},
  Z.\ Phys.\ C {\bf 75} (1997) 215;
%
Phys.\ Lett.\ B {\bf 437} (1998) 432;
ZEUS Collaboration: S.~Chekanov {\it et al.},  
Eur.\ Phys.\ J.\ C {\bf 24} (2002) 345.




\bibitem{two-channel}
V.~A.~Khoze, A.~D.~Martin and M.~G.~Ryskin,
Eur.\ Phys.\ J.\ C {\bf 18} (2000) 167;
 V.~A. Khoze, A.~D. Martin, M.~G. Ryskin and W.~J. Stirling,
Eur.\ Phys.\ J.\ C \textbf{35} (2004) 211.

\bibitem {KMR-phot} 
  V.~A.~Khoze, A.~D.~Martin and M.~G.~Ryskin,
  Eur.\ Phys.\ J.\ C {\bf 24} (2002) 459.

\bibitem {Ewerz-coupling} H.~G. Dosch, C.~Ewerz and V.~Schatz, 
Eur.\ Phys.\ J.\ C \textbf{24} (2002) 561.








\bibitem{alice} R. Schicker, see talk at this conference.




\end{thebibliography}

}
\end{footnotesize}
\end{document}